\documentclass[prd,showpacs,preprintnumbers]{revtex4} 
\def\sqr#1#2{{\vcenter{\vbox{\hrule height.#2pt\hbox{\vrule
width.#2pt height#1pt \kern#1pt\vrule width.#2pt}\hrule height.#2pt}}}}
 
\usepackage{latexsym}
\usepackage{bm}
\usepackage{graphicx}
\usepackage{color}

\begin{document}
\preprint{DCPT-10/29, KUNS-2284}

\title{Stability of Holographic Superconductors}

\author{Sugumi Kanno$^{1)}$}
\author{Jiro Soda$^{2)}$}
\affiliation{1) Centre for Particle Theory, Department of Mathematical 
Sciences, Durham University, Science Laboratories, South Road, Durham, 
DH1 3LE, United Kingdom}
\affiliation{2) Department of Physics,  Kyoto University, Kyoto, 606-8501, Japan
}

\date{\today}

\begin{abstract}
We study the dynamical stability of holographic superconductors.
We first classify perturbations around black hole background solutions
into vector and scalar sectors by means of a 2-dimensional
rotational symmetry. We prove the stability of the vector sector
by explicitly constructing the positive definite Hamiltonian. 
To reveal a mechanism for the stabilization of a superconducting phase,
we construct a quadratic action for the scalar sector.
From the action, we see the stability of black holes near a critical
point is determined by the equation of motion for a charged scalar field.
 We show the effective mass of the charged scalar field in hairy black holes
is always above the Breitenlohner-Freedman bound near the critical point
due to the backreaction of a gauge field.  
It implies the stability of the superconducting phase.
We also argue that the stability continues away from the critical point.
\end{abstract}

\maketitle

\section{Introduction}

It is believed that the AdS/CFT correspondence is useful to study
strongly coupled systems~\cite{Maldacena:1997re}. 
Remarkably, the AdS/CFT correspondence has been extended to the 
correspondence between gravity and condensed matter physics~\cite{Hartnoll:2009sz}. 
In particular, it has been shown that there exists a gravity dual
of a superconductor~\cite{Hartnoll:2008vx,Horowitz:2008bn,Hartnoll:2008}. 
Interestingly, the mean field
square root behavior of the Landau-Ginzburg
second order phase transition has been reproduced through the gravity dual 
of the superconductor. Moreover, it turned out that conductivity
has similar features to that of the
superconductor (see review articles and references 
therein~\cite{Herzog:2009xv,Horowitz:2010gk}). 

It is well known that the superconductor can be explained as 
the second order phase transition phenomena.
In terms of the Landau-Ginzburg theory, below a critical temperature,
the effective mass of the order parameter field
 becomes tachyonic and destabilize normal phase.
Eventually, a superconducting phase is realized as the new phase
where the stability is guaranteed by a quartic potential.
In the holographic description of the superconductor, the onset of the 
instability is well understood. The point is that  the coupling of a 
charged scalar field to a gauge field through covariant derivatives 
induces an effective mass term for 
the scalar field~\cite{Gubser:2008px}.
This term becomes relevant as temperature gets lowered at fixed charge
density, and eventually it makes the effective mass of the charged 
scalar field below the Breitenlohner-Freedman (BF) 
bound~\cite{Breitenlohner:1982jf},
and hence destabilize the system. 
 On the other hand, the fate of the instability is not fully understood.
In fact, in the gravity dual of superconductors, no quartic potential 
to stabilize the system after the instability is present.
Hence, although a trigger of the phase transition is clear,
a mechanism of stabilization of the superconducting phase is 
not apparent. In this paper, we study the dynamical stability
of the superconducting phase in the gravity dual model of the superconductor
and reveal that the superconductor is stabilized through a
backreaction of the gauge field. 

From the gravity point of view, the normal phase corresponds to
charged black holes in anti-de Sitter spacetime,
i.e., the Reissner-Nortstr\"{o}m-AdS black holes. The instability
below the critical temperature drives the black hole into hairy
black hole which corresponds to the superconducting phase. Hence, 
what we would like to prove is the stability of this hairy black hole.
Therefore, we can use the standard technique in the black hole perturbation
theory~\cite{Regge:1957td,Zerilli:1970se,Moncrief:1974am}. 
We classify general perturbations around the hairy black hole
into vector and scalar sectors by means of a 2-dimensional rotational 
symmetry of the black hole. We show the vector sector contains no 
unstable mode.
In order to complete the proof of stability, we examine the scalar sector.
By looking at the vicinity of the transition point, we demonstrate
how the stability is realized in the superconducting phase.
We also argue that the stability of the system 
persists away from the critical point.

The organization of this paper is as follows.
In section 2, we introduce the model and present background equations of
motion. Here, we describe the phase transition.
In section 3, we explain the Arnowitt-Deser-Misner (ADM) formalism which is useful to
perform efficient calculations and the stability analysis.
In section 4, we prove the stability of the vector sector by
constructing the positive definite Hamiltonian. 
In section 5, we reveal a mechanism for the stabilization of the 
superconducting phase through the analysis of the scalar sector.
Although the actual analysis is performed in the vicinity of the critical point,
we argue the stability holds even away from the critical point. 
The final section is devoted to the conclusion. 

\section{Gravity dual of superconductors}

In this section, we review the gravity/superconductor correspondence
\cite{Hartnoll:2008vx, Hartnoll:2008}.
Here, we take into account backreaction for completeness.

The action is given by
\begin{eqnarray}
S=\int d^4x \sqrt{-g}
\left[~\frac{1}{2\kappa^2}\left(
{\cal R} +\frac{6}{L^2}\right)
-\frac{1}{4}F^{\mu\nu}F_{\mu\nu}
-|\nabla\psi - iqA\psi |^2 - V(\psi)
~\right]  \ ,
\end{eqnarray}
where ${\cal R}$ is the 4-dimensional Ricci scalar 
and $L$ denotes the AdS curvature scale.
Here, we have incorporated the charged scalar field $\psi$ and the gauge
 field $A_\mu$ from which  we can calculate the field strength 
$F_{\mu\nu} = \partial_\mu A_\nu - \partial_\nu A_\mu$. 
As for the potential $V(\psi)$, we take the mass term, $V(\psi)= m^2 |\psi|^2 $.
In this paper, we assume that the mass is always above the BF bound.
Note that the coupling constant $q$ controls the strength of the 
backreaction.

Let us consider the static background. Then, the metric is given by:
\begin{eqnarray*}
ds^2_b=-f(r)e^{-\chi(r)} dt^2 + \frac{dr^2}{f(r)}
+ \frac{r^2}{L^2}\delta_{ab}dy^ady^b \ ,
\end{eqnarray*}
where $a, b$ denote $x, y$ coordinates.
The other fields are expressed by:
\begin{eqnarray}
A_\mu=(\phi(r),0,0,0)\,, \hspace{1cm}\psi =\psi(r) \ .
\end{eqnarray}
Note that the scalar field is taken to be real using the $U(1)$ gauge transformation.
Then, $tt$ and $rr$ components of the background Einstein equations yield
\begin{eqnarray}
&&f^\prime+\frac{1}{r}f-\frac{3r}{L^2}
+\kappa^2 r\left[
\frac{e^\chi}{2}\phi^{\prime2}
+m^2\psi^2
+f\left(
\psi^{\prime 2}+\frac{q^2\phi^2\psi^2e^\chi}{f^2}
\right)
\right]=0 \ ,
\label{tt}\\
&&\chi^{\prime}+2\kappa^2r \left(
\psi^{\prime2}
+\frac{q^2\phi^2\psi^2e^\chi}{f^2}
\right)=0  \ ,
\label{rr}
\end{eqnarray}
Here, a prime denotes derivative with respect to $r$.
while the gauge and the scalar equations become
\begin{eqnarray}
&&\phi^{\prime\prime}+\left(
\frac{\chi^\prime}{2}+\frac{2}{r}\right)\phi^\prime
-\frac{2q^2\psi^2}{f}\phi=0  \ ,
\label{maxwell}\\
&&\psi^{\prime\prime}+\left(
\frac{f^\prime}{f}-\frac{\chi^\prime}{2}+\frac{2}{r}
\right)\psi^\prime
+\left(
\frac{q^2\phi^2e^\chi}{f^2}
-\frac{m^2}{f}
\right)\psi=0  \ .
\label{scalar}
\end{eqnarray}
From the last equation, we see the gauge potential $\phi$ induces the effective
negative mass squared for the charged scalar field. 
This acts as a trigger of the phase transition.

By solving the above equations with appropriate boundary conditions,
we obtain the asymptotic behavior
\begin{eqnarray}
  \phi &=& \mu - \frac{\rho}{r} + \cdots\,,  \\
  \psi &=& \frac{\langle {\cal O}_\Delta \rangle}{ r^\Delta } + \cdots \ ,
\end{eqnarray}
where $\mu$ and $\rho$ are interpreted as a chemical potential and charge 
density of the dual theory on the boundary. $\langle {\cal O}_\Delta \rangle$ represents the expectation value of the operator $O_\Delta$
dual to the charged scalar field $\psi$. The exponent $\Delta$ is determined by
the mass as $\Delta=3/2 + \sqrt{9+4m^2L^2}/2$. This is the dictionary of AdS/CFT correspondence. 
When Hawking temperature is above a critical temperature, $T > T_c $, 
the solution is given by the Reissner-Nortstr\"{o}m-AdS 
black holes
\begin{eqnarray}
   \chi=\psi=0 \ , \quad 
   f= \frac{r^2}{L^2} - \frac{1}{r} \left( 
   \frac{r_+^3}{L^2} + \frac{\kappa^2\rho^2}{2r_+}\right)
   + \frac{\kappa^2\rho^2}{2r^2}
   \ ,\quad 
   \phi = \rho \left( \frac{1}{r_+} - \frac{1}{r} \right) 
   \ ,
\end{eqnarray}
where $r_+$ represents the horizon radius.
For $T<T_c$, the Reissner-Nordstr\"{o}m AdS black 
hole solutions become unstable and
new hairy black holes will be firmed. There, the expectation value
$\langle {\cal O}_\Delta \rangle$ 
has a non-trivial value proportional to $\sqrt{T_c -T}$. 
This is the phase transition from a normal phase to a superconducting phase.

In subsequent sections, we will investigate the stability
of the hairy black holes corresponding to the superconducting phase.

\section{Uses of ADM Formalism}

Although there are some indirect evidences of the stability of holographic
superconductors~\cite{Maeda:2010hf,Murata:2010dx}, it is important to give a direct evidence.
To this end, we need to perform perturbative analysis of the system introduced 
in the previous section. 
In this section, we explain a useful method to obtain a quadratic action
for black hole perturbations~\cite{Takahashi:2010th}.

Since the background spacetime is static, it is useful to work in
the ADM formalism.
Let us take the parameterization for the metric 
\begin{eqnarray}
ds^2 = - N^2 dt^2 + h_{ij}(dx^i+N^idt)(dx^j+N^jdt)\,,
\end{eqnarray}
where $i=r,x,y$ denotes the spatial coordinates,
 $N$ is the lapse function, and $N^i$ is the 
shift vector. Under this metric parameterization,
the Einstein-Hilbert action can be written in terms of
 the spatial curvature $R$ and the extrinsic curvature $K_{ij}$
 as
\begin{eqnarray}
S_{\rm R}&=&\frac{1}{2\kappa^2}\int d^4x \sqrt{-g}\left[~
{\cal R} + \frac{6}{L^2}~\right]\nonumber\\
&=&\frac{1}{2\kappa^2}\int dtd^3x\sqrt{h}
\left[~
NR + \frac{1}{N}
\left(
E^{ij}E_{ij} - E^2 
\right)
+ N\frac{6}{L^2}
~\right]\,,
\label{SR}
\end{eqnarray}
where for convenience we used the symmetric tensor $E_{ij}$ 
instead of the extrinsic curvature, which is defined by

\begin{eqnarray}
E_{ij} = \frac{1}{2}
\left[
\dot{h}_{ij} - N_{i;j} - N_{j;i}
\right] = NK_{ij}  \ .
\end{eqnarray}
Here, a dot and a semicolon are a time derivative and a covariant
derivative with respect to $h_{ij}$, respectively.
Similarly, the action for the gauge field reads
\begin{eqnarray}
S_{\rm A}&=&\int d^4x\sqrt{-g}
\left[
-\frac{1}{4}F^{\mu\nu}F_{\mu\nu}
\right]\nonumber\\
&=&\int d^4x \sqrt{h}N
\left[~
\frac{1}{2N^2}h^{ij}F_{ti}F_{tj}
-\frac{N^i}{N^2}\left(
h^{j\ell}-\frac{N^jN^\ell}{N^2}\right)F_{tj}F_{i\ell}
-\frac{1}{4}\left(h^{ik}-\frac{N^iN^k}{N^2}\right)
\left(h^{j\ell}-\frac{N^jN^\ell}{N^2}\right)F_{ij}F_{k\ell}
~\right]  \,. \qquad 
\label{SA}
\end{eqnarray}
The action for the charged scalar field is also written as
\begin{eqnarray}
S_{\psi}&=&\int d^4x\sqrt{-g}
\left[~
-|\nabla\psi - iqA\psi |^2 - V(\psi)
~\right]\nonumber\\
&=&\int d^4x \sqrt{h}N 
\left[~
\frac{1}{N^2}\{ 
\dot{\psi}^* + iqA_t\psi^* - N^i
\left(
\psi^*_{,i}+iqA_i\psi^*
\right)
\}
\{ 
\dot{\psi} - iqA_t\psi - N^j
\left(
\psi_{,j}-iqA_j\psi
\right)
\}
\right.\nonumber\\
&&\left.\hspace{2.1cm}
-h^{ij}
\left(
\psi^*_{,i}+iqA_i\psi^*
\right)
\left(
\psi_{,j}-iqA_j\psi
\right)
-V(\psi)
~\right]\,.
\label{SPSI}
\end{eqnarray}

Now, it is easy to see why the ADM formalism is useful for our purpose. 
In the ADM formalism, the background metric reads
\begin{eqnarray*}
ds^2_b=-N^2(r)dt^2 + h_{ij}(r)dx^idx^j\,,
\end{eqnarray*}
where
\begin{eqnarray}
N^2(r)=f(r)e^{-\chi(r)}\,\,,\hspace{0.5cm}
h_{ij}(r)dx^idx^j=\frac{dr^2}{f(r)}+\frac{r^2}{L^2} \delta_{ab}dy^ady^b  \ .
\label{bg}
\end{eqnarray}
Apparently, there is no shift vector $N^i$ in the background metric.
Hence, we find that $E_{ij}$ vanishes for the background. 
And, the gauge field has only time component, so we have $F_{ij}=0$. 
Thus, some of the terms in Eqs.~(\ref{SR}), (\ref{SA}) and 
(\ref{SPSI}) are already the second order quantities.
In this way, the ADM formalism makes calculations for obtaining the quadratic action
easier. In order to further simplify the calculations,  
 we take the variation of the total action
$S=S_R+S_A+S_\psi$ with respect to the lapse function $N$
to yield the Hamiltonian constraint equation:
\begin{eqnarray}
\frac{1}{2\kappa^2}\left(R+\frac{6}{L^2}\right)
-\frac{1}{2N^2}h^{ij}F_{ti}F_{tj}
-\frac{1}{N^2}q^2A_t^2\psi^*\psi
-h^{ij}\psi^*_{,i}\psi_{,j}-V(\psi)
=0\,.
\end{eqnarray}
Substituting this equation into the total action, 
we can simplify the part proportional to $\left( \sqrt{h}N \right)^{(2)}$
as
\begin{eqnarray}
\int d^4x 
\left(
\sqrt{h}N
\right)^{(2)}
\left[~
\frac{1}{N^2}h^{ij}F_{ti}F_{tj}
+\frac{2}{N^2}q^2A_t^2\psi^*\psi
~\right]\,,
\end{eqnarray}
where index $(2)$ means the second order quantity.\\

Since there exists a 2-dimensional plane symmetry in the black hole background,
we can decompose any tensor into vector and scalar sectors 
by means of the 2-dimensional rotation symmetry. Those sectors are
decoupled in the linear equations. Hence, we will consider the vector
and the scalar sectors separately. 

\section{Vector Sector }

In this section, we will prove the stability of the vector sector of perturbations.
To this aim, we use the formalism explained in the previous section. 
Hereafter, we set $L=1$ and $\kappa^2 =1$.

Let us consider the metric perturbations on the background metric (\ref{bg}),
\begin{eqnarray}
ds^2 = ds^2_b + \delta g_{\mu\nu}dx^\mu dx^\nu\,.
\end{eqnarray}
The vector sector of metric perturbations is generally expressed by 
\begin{equation}
\delta g_{\mu\nu} =\left(
\begin{array}{cccc}
& 0~ &  ~0~ & ~\delta g_{ta} 
\vspace{2mm}\\\vspace{2mm}
& \ast  & ~0~ & ~\delta g_{ra}              
\\
& \ast & \ast & \delta g_{ab}      
\end{array}
\right) \ , \hspace{5mm} * {\rm ~is ~symmetric ~part} \ ,
\end{equation}
where $\delta g_{ta}$, $\delta g_{ra}$ and 
$\delta g_{ab} = 2\zeta_{(a|b)}$ 
satisfy the divergence free
condition: $\delta g_{ta}{}^{|a}=\delta g_{ra}{}^{|a}=\zeta_a{}^{|a}=0$.
Here, $|$ represents a partial derivative.
Using the gauge transformation $x^\mu \rightarrow x^\mu - \xi^\mu$ with
\begin{eqnarray}
\xi_\mu = ( 0, 0, \xi_a )\,,
\end{eqnarray}
where $\xi_a{}^{|a}=0$, those variables transform into
\begin{eqnarray}
\delta g_{ta} \rightarrow \delta g_{ta} + \dot{\xi}_a\,,
\hspace{1cm}
\delta g_{ra} \rightarrow \delta g_{ra} + \xi_a^\prime -\frac{2}{r}\xi_a\,,
\hspace{1cm}
\delta g_{ab} \rightarrow \delta g_{ab} + 2\xi_{(a|b)}\,.
\end{eqnarray}
 Thus we can make $\delta g_{ab}$ vanish by choosing $\xi_a = -\zeta_a$
and we get the complete gauge fixing. This is called the Regge-Wheeler gauge.
In the Regge-Wheeler gauge, the perturbations that belong to the vector
sector can be written as
\begin{equation}
\delta g_{\mu\nu} =\left(
\begin{array}{cccc}
& 0~ &  ~0~ & ~v_a 
\vspace{2mm}\\\vspace{2mm}
& \ast  & ~0~ & ~w_a              
\\
& \ast & \ast & ~0      
\end{array}
\right) \ . 
\end{equation}
where $v_a{}^{|a}=w_a{}^{|a}=0$.
In the ADM formalism, this corresponds to
\begin{eqnarray}
 \delta N=0 \ , \quad \delta N_i = v_i \ , \quad 
 \delta h_{ij} =\left(
\begin{array}{ccc} 
&  0~ & ~w_a              
\vspace{2mm}\\
& \ast & ~0      
\end{array}
\right) \ . 
\end{eqnarray}
As to the gauge field, we can take
\begin{eqnarray}
\delta A_\mu = (0, 0, Z_a)\,,
\end{eqnarray}
where $Z_a$ satisfies $Z_a{}^{|a} =0$.

Now, we can calculate the quadratic action from Eqs.~(\ref{SR}), (\ref{SA}) and 
(\ref{SPSI}).
The quadratic part of Eq.~(\ref{SR}) is going to be
\begin{eqnarray}
S_R&=&\frac{1}{2}\int d^4x \left(
\sqrt{h}N
\right)^{(0)}
\left[~
R + \frac{1}{N^2}\left(
E^{ij}E_{ij}-E^2
\right)
~\right]^{(2)}
\nonumber\\
&=&
\frac{1}{2}\int d^4x r^2 e^{-\frac{\chi}{2}}
\left[~
- \frac{f^2\chi^\prime}{r^3} w^aw_a
+ \frac{f}{2r^4}w^a w_a{}^{|b}{}_{|b}
+ \frac{1}{2r^2}e^{\chi}
\left( \dot{w}_a - v^\prime_a + \frac{2}{r}v_a
\right)
\left( \dot{w}^a - v^{\prime a} + \frac{2}{r}v^a
\right)
\right.\nonumber\\
&&\hspace{3cm}\left.
- \frac{1}{2r^4f}e^{\chi}
v^av_a{}^{|b}{}_{|b}
~\right]\,.
\end{eqnarray} 
The quadratic part of Eq.~(\ref{SA}) becomes
\begin{eqnarray}
S_{\rm A}
&=&\int d^4x \left( 
\sqrt{h}N
\right)^{(0)}
\left[~
\frac{1}{2N^2}h^{ij}F_{ti}F_{tj}
-\frac{N^i}{N^2}h^{j\ell}F_{tj}F_{i\ell}
-\frac{1}{4}h^{ik}h^{j\ell}F_{ij}F_{k\ell}
~\right]^{(2)}
\nonumber\\
&& + \int d^4x \left(
\sqrt{h}N
\right)^{(2)}
\left[~
\frac{1}{N^2}h^{ij}F_{ti}F_{tj}
~\right]^{(0)}
\nonumber\\
&=& \int d^4x 
\left[~
e^{\frac{\chi}{2}}\phi^\prime w^a \dot{Z}_a
+ \frac{1}{2f}e^{\frac{\chi}{2}}\dot{Z}_a\dot{Z}^a
- e^{\frac{\chi}{2}}\phi^\prime v^a Z^\prime_a
- \frac{f}{2}e^{-\frac{\chi}{2}}Z^{\prime a}Z^\prime_a
- \frac{1}{2r^2}e^{-\frac{\chi}{2}}Z_{a|c}Z^{a|c}
~\right]\,.
\end{eqnarray}
The quadratic part of Eq.~(\ref{SPSI}) gives
\begin{eqnarray}
S_{\psi}
&=&\int d^4x \left(
\sqrt{h}N 
\right)^{(0)}
\left[~
\frac{1}{N^2}\{ 
\dot{\psi}^* + iqA_t\psi^* - N^i
\left(
\psi^*_{,i}+iqA_i\psi^*
\right)
\}
\{ 
\dot{\psi} - iqA_t\psi - N^j
\left(
\psi_{,j}-iqA_j\psi
\right)
\}
\right.\nonumber\\
&&\left.\hspace{3cm}
-h^{ij}
\left(
\psi^*_{,i}+iqA_i\psi^*
\right)
\left(
\psi_{,j}-iqA_j\psi
\right)
-V(\psi)
~\right]^{(2)}
\nonumber\\
&&
+ \int d^4x 
\left(
\sqrt{h}N
\right)^{(2)}
\left[~
\frac{2}{N^2}q^2A_t^2\psi^*\psi
~\right]^{(0)}
\nonumber\\
&=& \int d^4x 
\left[~
-2\frac{q^2}{f}e^{\frac{\chi}{2}}\phi\psi^2 v^aZ_a
-f^2e^{-\frac{\chi}{2}}\psi^{ \prime 2} w^a w_a
-q^2 e^{-\frac{\chi}{2}}\psi^2 Z^aZ_a
-q^2e^{\frac{\chi}{2}}\phi^2\psi^2 w^aw_a
~\right]\,.
\end{eqnarray}
The total quadratic action for the vector sector becomes
\begin{eqnarray}
S&=&S_{\rm R} + S_{\rm A} + S_{\psi}  \nonumber\\
&=& \frac{1}{2}
\int dt dr d^2 k e^{-\frac{\chi}{2}}
\left[~ 
 - \frac{\chi^\prime}{r}f^2 w^aw_a
- \frac{k_b^2 f}{2r^2}w^a w_a
+ \frac{e^\chi}{2}
\left( \dot{w}_a - v^\prime_a + \frac{2}{r}v_a
\right)
\left( \dot{w}^a - v^{\prime a} + \frac{2}{r}v^a
\right)
+ \frac{k_b^2}{2r^2f}e^\chi
 v^av_a
~\right]
\nonumber\\
&& +\int dt dr d^2 k 
\left[~
e^{\frac{\chi}{2}}\phi^\prime w^a \dot{Z}_a
+ \frac{1}{2f}e^{\frac{\chi}{2}}\dot{Z}_a\dot{Z}^a
- e^{\frac{\chi}{2}}\phi^\prime v^a Z^\prime_a
- \frac{f}{2}e^{-\frac{\chi}{2}}Z^{\prime a}Z^\prime_a
- \frac{k_b^2}{2r^2}e^{-\frac{\chi}{2}}Z_aZ^a
\right.\nonumber\\
&&\hspace{2.3cm}\left.
-2\frac{q^2}{f}e^{\frac{\chi}{2}}\phi\psi^2 v^aZ_a
-f^2e^{-\frac{\chi}{2}}\psi^{\prime 2} w^a w_a
-q^2 e^{-\frac{\chi}{2}}\psi^2 Z^aZ_a
-q^2e^{\frac{\chi}{2}}\phi^2\psi^2 w^a w_a
~\right]\,,
\label{action1}
\end{eqnarray}
where we moved on to the Fourier space with respect to $x,y$ coordinates.
Note that there are three unknown variable $v_a, w_a$ and $Z_a$,
 which satisfy the transverse conditions. Among them, $v_a$ is
 not a dynamical one which should be eliminated.

The action Eq.~(\ref{action1}) becomes,
\begin{eqnarray}
S
&=& \int dt dr d^2 k e^{-\frac{\chi}{2}}
\left[~ 
- \frac{k_b^2 f}{4r^2}w^a w_a
+ \frac{e^\chi}{4}
\left( \dot{w}_a - v^\prime_a + \frac{2}{r}v_a
\right)
\left( \dot{w}^a - v^{\prime a} + \frac{2}{r}v^a
\right)
+ \frac{k_b^2}{4r^2f}e^\chi
v^av_a
+ e^\chi\phi^\prime w^a \dot{Z}_a
\right.\nonumber\\
&& \left.\hspace{2.7cm}
+ \frac{1}{2f}e^\chi\dot{Z}_a\dot{Z}^a
- e^\chi\phi^\prime v^a Z^\prime_a
- \frac{f}{2}Z^{\prime a}Z^\prime_a
- \frac{k_b^2}{2r^2}Z_aZ^a
-2\frac{q^2}{f}e^\chi\phi\psi^2 v^aZ_a
-q^2 \psi^2 Z^aZ_a
~\right]\,,
\end{eqnarray}
where we eliminated some terms proportional to $w_aw^a$ 
 using the background equation (\ref{rr}). Now we want to eliminate the non-dynamical
field $v^a$ by completing the square with respect to $v_a$.
In doing so, we need the Hamiltonian formalism in order to
eliminate $v_a^\prime v^{\prime a}$. Defining the conjugate momentum
\begin{eqnarray}
P_w^a&=&\frac{\partial\cal{L}}{\partial\dot{w}_a}
=\frac{1}{2}e^{\frac{\chi}{2}}
\left(
\dot{w}^a-v^{\prime a}+\frac{2}{r}v^a
\right)   \ , \\
P_z^a&=&\frac{\partial\cal{L}}{\partial\dot{Z}_a}
=e^{\frac{\chi}{2}}
\left(\phi^\prime w^a
+\frac{1}{f}\dot{Z}^a
\right)   \ ,
\end{eqnarray}
we obtain the Hamiltonian
\begin{eqnarray}
H&=& \int drd^2k_a\left[~
P_w^a\dot{w}_a+P_z^a\dot{Z}_a-{\cal L}
~\right]
\nonumber\\
&=&\int drd^2k_a \left[~
e^{-\frac{\chi}{2}}P_w^aP_{wa}
+\left(
v_a^\prime-\frac{2}{r}v_a
\right)P_w^a
+\frac{f}{2}e^{-\frac{\chi}{2}}\left(
P_z^a-e^{\frac{\chi}{2}}\phi^\prime w^a
\right)\left(
P_{za}-e^{\frac{\chi}{2}}\phi^\prime w_a
\right)
\right.\nonumber\\
&&\left.\hspace{1.7cm}
+\frac{k_b^2f}{4r^2}e^{-\frac{\chi}{2}}w^aw_a
-\frac{k_b^2}{4r^2f}e^{\frac{\chi}{2}}v^av_a
+e^{\frac{\chi}{2}}\phi^\prime v^aZ_a^\prime
+\frac{f}{2}e^{-\frac{\chi}{2}}Z_a^\prime Z^{\prime a}
\right.\nonumber\\
&&\left.\hspace{1.7cm}
+\frac{k_c^2}{2r^2}e^{-\frac{\chi}{2}}Z_aZ^a
+2\frac{q^2}{f}e^{\frac{\chi}{2}}\phi\psi^2 v^aZ_a
+q^2e^{-\frac{\chi}{2}}\psi^2 Z^aZ_a
~\right]         \ .
\end{eqnarray}
Let us see the terms containing $v_a$. After integrating by parts
for $v_a^\prime$, we have
\begin{eqnarray}
-v_a\left(
P_w^{a\prime}+\frac{2}{r}P_w^a-e^{\frac{\chi}{2}}\phi^\prime Z_a^\prime
-2\frac{q^2}{f}e^{\frac{\chi}{2}}\phi\psi^2 Z^a
\right)
-\frac{k_b^2}{4r^2f}e^{\frac{\chi}{2}}v^av_a\,.
\end{eqnarray}
By completing the square for the second and third terms as
\begin{eqnarray}
&&-\frac{1}{4r^2f}e^{\frac{\chi}{2}}k_b^2
\left\{~
v^a+\frac{2r^2f}{k_b^2}e^{-\frac{\chi}{2}}
\left(
P_w^{a\prime}+\frac{2}{r}P_w^a-e^{\frac{\chi}{2}}\phi^\prime Z_a^\prime
-2\frac{q^2}{f}e^{\frac{\chi}{2}}\phi\psi^2 Z^a
\right)~\right\}^2
\nonumber\\
&&\hspace{1cm}
+\frac{r^2f}{k_b^2}e^{-\frac{\chi}{2}}
\left\{~
P_w^{a\prime}+\frac{2}{r}P_w^a-e^{\frac{\chi}{2}}\phi^\prime Z_a^\prime
-2\frac{q^2}{f}e^{\frac{\chi}{2}}\phi\psi^2 Z^a
~\right\}^2\,.
\end{eqnarray}
the variable $v_a$ can be eliminated because the first squared term vanishes
after substituting the equation of motion for $v_a$. Then the Hamiltonian
becomes
\begin{eqnarray}
H&=&
\int drd^2k
\left[~
\frac{r^2f}{k_b^2}e^{-\frac{\chi}{2}}
\left\{~
P_w^{a\prime}+\frac{2}{r}P_w^a-e^{\frac{\chi}{2}}\phi^\prime Z_a^\prime
-2\frac{q^2}{f}e^{\frac{\chi}{2}}\phi\psi^2 Z^a
~\right\}^2
\right.\nonumber\\
&&\left.\hspace{1.7cm}
+ e^{-\frac{\chi}{2}}P_w^aP_{wa}
+\frac{f}{2}e^{-\frac{\chi}{2}}\left(
P_z^a-e^{\frac{\chi}{2}}\phi^\prime w^a
\right)\left(
P_{za}-e^{\frac{\chi}{2}}\phi^\prime w_a
\right)
+\frac{k_b^2f}{4r^2}e^{-\frac{\chi}{2}}w^aw_a
\right.\nonumber\\
&&\left.\hspace{1.7cm}
+\frac{f}{2}e^{-\frac{\chi}{2}}Z_a^\prime Z^{\prime a}
+\frac{k_c^2}{2r^2}e^{-\frac{\chi}{2}}Z_aZ^a
+q^2e^{-\frac{\chi}{2}}\psi^2 Z^aZ_a
~\right]\,.
\end{eqnarray}
We find the Hamiltonian consists of positive terms. 
This implies the stability of the system. The argument is as follows.
Suppose an unstable mode exists, then the Hamiltonian must grow.
However, since the Lagrangian
has a time translation invariance, the energy of the system is conserved.
This contradicts the growth of the Hamiltonian.
Therefore, the unstable mode cannot exist in the vector sector. 
Note that this is true either for the normal phase or the superconducting phase.

\section{Scalar Sector}

In the previous section, we have proved the stability of the vector sector
for both the normal and superconducting phase.
Thus, if unstable modes exist, they must belong to the scalar sector.
Above the critical temperature, the system is the Reissner-Nordstr\"{o}m-AdS
black hole with a trivial scalar field. 
For this case, we know the system is stable~\cite{Chandra}.
Below the critical temperature, the effective mass of the charged scalar field
violates BF bound, then the system ceases to be stable.
Eventually, a hairy black hole is formed. It is believed that
the new hairy black hole is stable. In this section, we reveal
the stabilization mechanism of the hairy black hole.

The scalar sector of metric perturbations is 
generally expressed by seven
degrees of freedom
\begin{equation}
\delta g_{\mu\nu} =\left(
\begin{array}{cccc}
& \delta g_{tt}~ &  ~\delta g_{tr}~ & ~\delta g_{ta} 
\vspace{2mm}\\\vspace{2mm}
& \ast  & ~\delta g_{rr}~ & ~\delta g_{ra}              
\\
& \ast & \ast & \delta g_{ab}      
\end{array}
\right) \ , \hspace{5mm} * {\rm ~is ~symmetric ~part} \ ,
\end{equation}
where $\delta g_{ab} = \zeta_1 \delta_{ab}+ \zeta_{2|ab} $
 has two degrees of freedom in the scalar sector.
Using the gauge transformation with
\begin{eqnarray}
\xi_\mu = ( \xi_t,~\xi_r,~\xi_{|a} )  \ ,
\end{eqnarray}
metric perturbations transform as
\begin{eqnarray}
&&\delta g_{tt} \rightarrow \delta g_{tt} + 2\dot{\xi}_t 
- f^2e^{-\chi}\left(\frac{f^\prime}{f}-\chi^\prime\right)\xi_r\,,
\nonumber\\
&&\delta g_{tr} \rightarrow \delta g_{tr} + \xi_t^\prime +\dot{\xi}_r
-\left(\frac{f^\prime}{f}-\chi^\prime\right)\xi_t\,,
\nonumber\\
&&\delta g_{ta} \rightarrow \delta g_{ta} + \xi_{t|a}
+ \dot{\xi}_{|a}\,,
\nonumber\\
&&\delta g_{rr} \rightarrow \delta g_{rr} + 2\xi_r^\prime
+\frac{f^\prime}{f}\xi_r\,,
\nonumber\\
&&\delta g_{ra} \rightarrow \delta g_{ra} + \xi_{r|a} 
+\xi_{|a}^\prime -\frac{2}{r}\xi_{|a}\,,
\nonumber\\
&&\delta g_{ab} \rightarrow \delta g_{ab} + 2\xi_{|ab}
+ 2fr \xi_r \delta_{ab}  \ .
\end{eqnarray}
Using the gauge degree of freedom of $\xi_t$, we can eliminate 
$\delta g_{ta}$. 
Using the gauge degree of freedom of $\xi$ and $\xi_r$, 
we can take $\delta g_{ab}=0$. 
Therefore, we take the following gauge.
\begin{equation}
\delta g_{\mu\nu} =\left(
\begin{array}{cccc}
& N^2\bar{H}~ &  ~H_1~ & 0 
\vspace{2mm}\\\vspace{2mm}
& \ast  & ~H/f~ & ~w_{|a}              
\\
& \ast & \ast & 0      
\end{array}
\right) \ . \hspace{5mm} * {\rm ~is ~symmetric ~part}.
\end{equation}
where the components of metric perturbations $w, H,H_1$ 
and $\bar{H}$ depend on $(t,r,a)$.
In the ADM formalism, we can put
\begin{eqnarray}
 \delta N= N \left( \sqrt{1- \bar{H}} -1 \right) \ , \quad \delta N_r = H_1 \ , \quad 
 \delta h_{ij} =\left(
\begin{array}{ccc} 
& H/f~ & ~w_{|a}              
\vspace{2mm}\\
& \ast & ~0      
\end{array}
\right) \ . 
\end{eqnarray}
In addition to the metric perturbations, we need to consider
fluctuations of the gauge field $\delta A_\mu$ 
and the charged scalar field $\delta \psi$, $\delta \psi^*$.
 As to the gauge field, we can take
\begin{eqnarray}
\delta A_\mu = (~\lambda, ~\alpha, ~\beta_{|a} ~)\,.
\end{eqnarray}
Using the $U(1)$ gauge transformation, we can take the scalar field
perturbations to be real $\delta \psi^* = \delta \psi$. 
However, for the trivial background $\psi=0$, this gauge fixing is
singular because the phase loses its meaning when the amplitude is zero.

Now, we can calculate the quadratic action for perturbed quantities
using the ADM formalism. 
The action for the gravity is calculated as 
\begin{eqnarray}
S_{\rm R} 
&=& \frac{1}{2}\int d^4x \left[
\left(
\sqrt{h}N
\right)^{(0)}R^{(2)}
+ \left(
\sqrt{h}N
\right)^{(1)}R^{(1)}
+ \left( \frac{\sqrt{h}}{N}\right)^{(0)}
\left(
E^{ij}E_{ij} - E^2
\right)^{(2)}
\right]
\nonumber\\
&=& \frac{1}{2}\int d^4x e^{-\frac{\chi}{2}}
\left[~ 
-\frac{1}{2}r^2\bar{H}
\left(
2\frac{f}{r}H^\prime-\frac{1}{r^2}H^{|a}{}_{|a}
+ 2\frac{f}{r^2}H + 2\frac{f^\prime}{r}H
+ 2\frac{f}{r^2}w^{\prime |a}{}_{|a} 
+ \frac{f^\prime}{r^2}w^{|a}{}_{|a}
+ 2\frac{f}{r^3}w^{|a}{}_{|a}
\right)
\right.\nonumber\\
&&\hspace{2.3cm}\left.
+\frac{rf}{2}\left(
\frac{1}{r}+\frac{f^\prime}{f}-\frac{3}{2}\chi^\prime
\right)H^2
-\frac{f}{2}\left(
\frac{f^\prime}{f}+\frac{2}{r}-\chi^\prime
\right)H^{|a}w_{|a}
-\frac{f^2\chi^\prime}{r}w^{|a}w_{|a}
\right.\nonumber\\
&&\hspace{2.3cm}\left.
+ \frac{e^\chi}{2}
\left( \dot{w} - H_1 \right)_{|a}
\left( \dot{w} -H_1 \right)^{|a}
+2re^\chi H_1\dot{H}
+rf\chi^\prime e^\chi H_1^2
~\right]\,.
\end{eqnarray}
In vacuum case, $\chi$ vanishes and 
the above action gives the Zerilli equation~\cite{Zerilli:1970se}.
In the present case, we have other fields.
The action for the gauge field becomes
\begin{eqnarray}
S_A&=&
\int d^4x \left[~
\left( 
\sqrt{h}N
\right)^{(0)}
\left\{~
\frac{1}{2N^2}h^{ij}F_{ti}F_{tj}
-\frac{N^i}{N^2}h^{j\ell}F_{tj}F_{i\ell}
-\frac{1}{4}h^{ik}h^{j\ell}F_{ij}F_{k\ell}
~\right\}^{(2)}
\right.\nonumber\\
&& \left.\hspace{1.4cm}
+ \left(
\sqrt{h}N
\right)^{(1)}
\left\{~
\frac{1}{2N^2}h^{ij}F_{ti}F_{tj}
~\right\}^{(1)}
+ \left(
\sqrt{h}N
\right)^{(2)}
\left\{~
\frac{1}{N^2}h^{ij}F_{ti}F_{tj}
~\right\}^{(0)}
~\right]
\nonumber\\
&=& \int d^4x r^2 e^{\frac{\chi}{2}}
\left[~
\frac{1}{2}\left\{
\dot{\alpha} - \lambda^\prime + \frac{1}{2}\phi^\prime
\left( H - \bar{H} \right)
\right\}^2
+ \frac{1}{r^2}\phi^\prime w^{|a} 
\left( \dot{\beta} - \lambda \right)_{|a}
+ r^2\phi^\prime\gamma^{|a}\left( \alpha - \beta^\prime \right)_{|a}
\right.\nonumber\\
&&\hspace{2cm}\left.
+ \frac{1}{2r^2 f}\left( \dot{\beta} - \lambda \right)_{|a}
\left( \dot{\beta} - \lambda \right)^{|a}
-\frac{f}{2r^2}e^{-\chi}\left( \beta^\prime - \alpha \right)_{|a}
\left( \beta^\prime - \alpha \right)^{|a}
~\right]\,.
\end{eqnarray}
Since we have eliminated the phase of the charged scalar field, the corresponding 
physical degree is absorbed by the gauge field. Then, 
$\alpha$ and $\beta$ are dynamical degrees describing the massive gauge field.
The time component of the perturbed vector $\lambda$ is a Lagrange multiplier
and hence not dynamical.
Finally, the action for the scalar field is given by
\begin{eqnarray}
S_\psi 
&=&\int d^4x \left(
\sqrt{h}N 
\right)^{(0)}
\left[~
\frac{1}{N^2}\{ 
\dot{\psi} + iqA_t\psi - N^i
\left(
\psi_{,i}+iqA_i\psi
\right)
\}
\{ 
\dot{\psi} - iqA_t\psi - N^j
\left(
\psi_{,j}-iqA_j\psi
\right)
\}
\right.\nonumber\\
&&\left.\hspace{3cm}
-h^{ij}
\left(
\psi_{,i}+iqA_i\psi
\right)
\left(
\psi_{,j}-iqA_j\psi
\right)
-V(\psi)
~\right]^{(2)}
\nonumber\\
&&
+ \int d^4x \left(
\sqrt{h}N 
\right)^{(1)}
\left[~
\frac{1}{N^2}\{ 
\dot{\psi} + iqA_t\psi - N^i
\left(
\psi_{,i}+iqA_i\psi
\right)
\}
\{ 
\dot{\psi} - iqA_t\psi - N^j
\left(
\psi_{,j}-iqA_j\psi
\right)
\}
\right.\nonumber\\
&&\left.\hspace{3.3cm}
-h^{ij}
\left(
\psi_{,i}+iqA_i\psi
\right)
\left(
\psi_{,j}-iqA_j\psi
\right)
-V(\psi)
~\right]^{(1)}
\nonumber\\
&&
+ \int d^4x 
\left(
\sqrt{h}N
\right)^{(2)}
\left[~
\frac{2}{N^2}q^2A_t^2\psi^2
~\right]^{(0)}
\nonumber\\
&=& \int d^4x~ r^2 e^{-\frac{\chi}{2}}
\left[~  \frac{e^\chi}{f}\left(
\delta\dot{\psi}^2
- 2 f\psi^\prime H_1 \delta\dot{\psi}
+ f^2 \psi^{\prime 2} H_1^2
\right)
- 2 q^2 e^\chi\phi \psi^2 H_1 \alpha
+ q^2\phi^2 \delta\psi^2  
+ q^2\psi^2 \lambda^2
\right.\nonumber\\
&&\hspace{2.4cm}\left.
+ 4q^2\phi \psi \delta\psi \lambda
-q^2 f \psi^2 \alpha^2 
-\frac{q^2}{r^2}\psi^2 \beta^{|a}\beta_{|a}
- f\delta\psi^{\prime 2} 
- \frac{1}{r^2}\delta\psi_{|a}\delta\psi^{|a}
\right.\nonumber\\
&&\hspace{2.4cm}\left.
- f\psi^{\prime 2}H^2 - \frac{f^2}{r^2}\psi^{\prime 2}
w^{|a}w_{|a}
+ 2 f\psi^{\prime } H \delta\psi^\prime 
+ 2 \frac{f}{r^2}\psi^\prime w^{|a} \delta\psi_{|a}
-  m^2 \delta\psi^2
~\right]
\nonumber\\
&&
+ \int d^4x~ \frac{1}{2}r^2e^{-\frac{\chi}{2}}\left( H - \bar{H} \right)
\left[~
2q^2\phi\psi^2 \lambda
+ 2 q^2\phi^2  \psi \delta\psi
+f\psi^{\prime 2} H 
- 2 f \psi^{\prime } \delta\psi^\prime
- 2 m^2 \psi \delta\psi
~\right]
\nonumber\\
&&
+ \int d^4x~ 2\frac{q^2r^2}{f}e^{\frac{\chi}{2}}\phi^2 \psi^2 
\left[~
-\frac{1}{8}\bar{H}^2 - \frac{1}{8}H^2
-\frac{f}{2r^2}w^{|a}w_{|a}   -\frac{1}{4}\bar{H}H
~\right]   \,.
\end{eqnarray}
Note that we have eight unknown variables
$
\bar{H},~H_1,~H,~w,~\lambda,~\alpha,~\beta,~\delta\psi
$. Among these, only four variables are physical. 
Thus, the total action in Fourier space becomes
\begin{eqnarray}
S 
&=& \frac{1}{2}\int dt dr d^2 k 
\left[~ 
-\frac{1}{2}r^2e^{-\frac{\chi}{2}}\bar{H}
\left(
2\frac{f}{r}H^\prime + \frac{k_a^{2}}{r^2}H
+ 2\frac{f}{r^2}H + 2\frac{f^\prime}{r}H
- 2\frac{k_a^2 f}{r^2}w^{\prime} 
- \frac{k_a^2 f^\prime}{r^2}w
- 2\frac{k_a^2 f}{r^3}w
\right)
\right.\nonumber\\
&&\hspace{2.5cm}\left.
+\frac{rf}{2}\left(
\frac{1}{r}+\frac{f^\prime}{f}-\frac{3}{2}\chi^\prime
\right)e^{-\frac{\chi}{2}}H^2
-\frac{k_a^2 f}{2}\left(
\frac{f^\prime}{f}+\frac{2}{r}-\chi^\prime
\right)e^{-\frac{\chi}{2}}Hw
\right.\nonumber\\
&&\hspace{2.5cm}\left.
+ \frac{k_a^2}{2}e^{\frac{\chi}{2}}
\left( \dot{w} - H_1
\right)^2
+2r e^{\frac{\chi}{2}} H_1\dot{H}
+rf\chi^\prime e^{\frac{\chi}{2}} H_1^2
~\right]
\nonumber\\
&&
+\int dt dr d^2k r^2 e^{\frac{\chi}{2}}
\left[~
\frac{1}{2}\left\{
\dot{\alpha} - \lambda^\prime + \frac{1}{2}\phi^\prime
\left( H - \bar{H} \right)
\right\}^2
+ \frac{k_a^2}{r^2}\phi^\prime w 
\left( \dot{\beta} - \lambda \right)
+ \frac{k_a^2}{2r^2 f}\left( \dot{\beta} - \lambda \right)^2
-\frac{k_a^2f}{2r^2}e^{-\chi}\left( \beta^\prime - \alpha \right)^2
~\right]
\nonumber\\
&&
+ \int dt dr d^2k r^2 e^{-\frac{\chi}{2}}
\left[~
\frac{e^\chi}{f}\left(
\delta\dot{\psi}-f\psi^\prime H_1
\right)^2
-2q^2e^\chi\phi\psi^2 H_1\alpha + q^2\phi^2\delta\psi^2
+q^2\psi^2\lambda^2 + 4q^2\phi\psi\delta\psi \lambda
-q^2f\psi^2\alpha^2
\right.\nonumber\\
&&\hspace{3.2cm}\left.
-\frac{k_a^2 q^2}{r^2}\psi^2\beta^2
-f\delta\psi^{\prime 2} 
- \frac{k_a^2}{r^2}\delta\psi^2
-f\psi^{\prime 2}H^2 + 2f\psi^\prime\delta\psi^\prime H
+2\frac{k_a^2 f}{r^2}\psi^\prime w \delta\psi
-  m^2 \delta\psi^2
~\right]
\nonumber\\
&&
+ \int dt dr d^2k \frac{1}{2}r^2 e^{-\frac{\chi}{2}}
\left( H - \bar{H} \right)
\left[~
2q^2\phi\psi^2\lambda + 2q^2\phi^2\psi\delta\psi
+ f\psi^{\prime 2}H
-2f\psi^\prime \delta\psi^\prime
- 2 m^2 \psi \delta\psi
~\right]
\nonumber\\
&&
+ \int dt dr d^2k r^2 e^{\frac{\chi}{2}}\frac{2q^2}{f}\phi^2\psi^2
\left[~
-\frac{1}{8}\bar{H}^2 - \frac{1}{8}H^2 -\frac{1}{4}\bar{H}H
~\right]\,,
\end{eqnarray}
where Eq.~(\ref{rr}) is used.

Now, it is convenient to use the Hamiltonian formalism
for getting an insight. 
Defining the canonical conjugate momentum:
\begin{eqnarray}
P_w&=&\frac{\partial\cal{L}}{\partial\dot{w}}
=\frac{k_a^2}{2}e^{\frac{\chi}{2}}
\left(
\dot{w} - H_1
\right)\,,\\
P_\alpha&=&\frac{\partial\cal{L}}{\partial\dot{\alpha}}
=r^2 e^{\frac{\chi}{2}}
\left\{
\dot{\alpha}
- \lambda^\prime + 
\frac{1}{2}\phi^\prime\left(
H - \bar{H}
\right)
\right\}\,,\\
P_\beta&=&\frac{\partial\cal{L}}{\partial\dot{\beta}}
=k_a^2e^{\frac{\chi}{2}}
\phi^\prime w
+\frac{k_a^2}{f}e^{\frac{\chi}{2}}\left(
\dot{\beta}-\lambda
\right)\,,\\
P_{\delta\psi}&=&\frac{\partial\cal{L}}{\partial\dot{\delta\psi}}
=2\frac{r^2}{f}e^{\frac{\chi}{2}}\left(
\delta\dot{\psi} - f\psi^\prime H_1
\right)\,,
\end{eqnarray}
we can perform the Legendre transformation and obtain
the Hamiltonian density 
\begin{eqnarray}
\cal{H}&=& \frac{1}{k_a^2}e^{-\frac{\chi}{2}} P_w^2 
+ \frac{1}{2r^2}e^{-\frac{\chi}{2}}P_\alpha^2
+ \frac{f}{2k_a^2}e^{-\frac{\chi}{2}}P_\beta^2
+ \frac{f}{4r^2}e^{-\frac{\chi}{2}}P_{\delta\psi}^2
\nonumber\\
&&
+ P_w H_1 + f\psi^\prime P_{\delta\psi} H_1
- \frac{1}{2}\phi^\prime P_\alpha\left(
H - \bar{H}
\right)
- f\phi^\prime P_\beta w 
+ P_\alpha \lambda^\prime 
+ P_\beta \lambda
\nonumber\\
&&
+ \frac{rf}{2}e^{-\frac{\chi}{2}}\bar{H}H^\prime
+ \frac{rf}{4}e^{-\frac{\chi}{2}}\left(
\frac{k_a^2}{rf}+\frac{2}{r}+2\frac{f^\prime}{f}
\right)\bar{H}H
- \frac{k_a^2 f}{2}e^{-\frac{\chi}{2}}\bar{H}w^\prime
-\frac{k_a^2 f}{4}e^{-\frac{\chi}{2}}\left(
\frac{f^\prime}{f}+\frac{2}{r}
\right)\bar{H}w
\nonumber\\
&&
-\frac{rf}{4}\left(
\frac{1}{r}+\frac{f^\prime}{f}-\frac{3}{2}\chi^\prime
\right)e^{-\frac{\chi}{2}}H^2
+r^2fe^{-\frac{\chi}{2}}\psi^{\prime 2}H^2
-\frac{r^2f}{2}e^{-\frac{\chi}{2}}\psi^{\prime 2}H\left(
H-\bar{H}
\right)
+\frac{q^2r^2}{4f}e^{\frac{\chi}{2}}\phi^2\psi^2\left(
H+\bar{H}
\right)^2
\nonumber\\
&&
+\frac{k_a^2 f}{4}\left(
\frac{f^\prime}{f}+\frac{2}{r}-\chi^\prime
\right)e^{-\frac{\chi}{2}}Hw
-re^{-\frac{\chi}{2}}H_1\dot{H}
-\frac{rf}{2}\chi^\prime e^{\frac{\chi}{2}}H_1^2
+\frac{k_a^2 f}{2}e^{\frac{\chi}{2}}\phi^{\prime 2}w^2
\nonumber\\
&&
+\frac{k_a^2 f}{2}e^{-\frac{\chi}{2}}\left(
\beta^\prime - \alpha
\right)^2
+2q^2r^2e^{\frac{\chi}{2}}\phi\psi^2H_1\alpha
+q^2r^2fe^{-\frac{\chi}{2}}\psi^2\alpha^2
+k_a^2 q^2 e^{-\frac{\chi}{2}} \psi^2 \beta^2
\nonumber\\
&&
- 2r^2fe^{-\frac{\chi}{2}}\psi^\prime H \delta\psi^\prime
- 2k_a^2fe^{-\frac{\chi}{2}}\psi^\prime w \delta\psi
- \frac{r^2}{2}e^{-\frac{\chi}{2}}\left(
H-\bar{H}
\right)\left(
2q^2\phi^2\psi\delta\psi - 2f\psi^\prime\delta\psi^\prime
- 2 m^2 \psi \delta\psi
\right)
\nonumber\\
&&
+ r^2fe^{-\frac{\chi}{2}}\delta\psi^{\prime 2}
+ k_a^2e^{-\frac{\chi}{2}}\delta\psi^2
+ r^2 e^{-\frac{\chi}{2}} m^2 \delta\psi^2
\nonumber\\
&&
- q^2r^2e^{-\frac{\chi}{2}}\phi^2\delta\psi^2
- q^2r^2e^{-\frac{\chi}{2}}\psi^2\left\{
\lambda^2 + 4\frac{\phi}{\psi}\delta\psi \lambda
+ \phi\left(
H-\bar{H}
\right)\lambda
\right\}\,.
\label{hamiltonian}
\end{eqnarray}

First, we consider the normal phase $\psi = 0$.
In this case, we find the scalar field perturbation 
decouples from the other parts in Eq.~(\ref{hamiltonian}) 
and reads
\begin{eqnarray}
{\cal H}_{\delta\psi} =
 \frac{f}{4r^2} e^{-\frac{\chi}{2}} P_{\delta\psi}^2
+ r^2fe^{-\frac{\chi}{2}}\delta\psi^{\prime 2}
+ k_a^2e^{-\frac{\chi}{2}}\delta\psi^2
+ r^2 e^{-\frac{\chi}{2}} m^2 \delta\psi^2
- q^2r^2e^{-\frac{\chi}{2}}\phi^2\delta\psi^2 \ .
\label{h:psi}
\end{eqnarray}
As the Reissner-Nordstr\"{o}m-AdS black hole is known to be stable, the instability of
the system should be caused by the scalar field. 
Apparently, the last term in Eq.~(\ref{h:psi}) could
destabilize the system. Indeed, as we lower the Hawking temperature, 
the potential $\phi\propto 1/r$
becomes larger near the horizon. Then, the effective mass 
$m^2_{\rm eff} = m^2 - q^2 \phi^2$ gets smaller.
Eventually, below the critical temperature, the effective mass
$m^2_{\rm eff} = m^2 - q^2 \phi^2$ violates the BF bound 
and the system becomes unstable.

Now, we see what happens in the superconducting phase $\psi \neq 0$.
In order to make the discussion clear, we focus on the stability of the 
region near the critical point where $\psi $ is close to zero
and the Reissner-Nordstr\"{o}m-AdS black hole becomes a good approximation.
The scalar perturbation $\delta \psi$ decouples from the metric perturbation
in the Reissner-Nordstr\"{o}m-AdS limit $\psi\rightarrow 0$. 
Hence, in the limit, we can set $\bar{H}=H_1=H=w=0$ when we look 
at the scalar field. In a sense, this is close to the probe limit.
Note that we have four variables $\alpha , \beta , \lambda , \delta \psi$
among which only $\lambda$ is unphysical. 
Now, the relevant part of the Hamiltonian is given by
\begin{eqnarray}
H&=& \int drd^2k_a \left[~
 \frac{1}{2r^2}e^{-\frac{\chi}{2}}P_\alpha^2
+ \frac{f}{2k_a^2}e^{-\frac{\chi}{2}}P_\beta^2
+ \frac{f}{4r^2}e^{-\frac{\chi}{2}}P_{\delta\psi}^2
+\frac{k_a^2 f}{2}e^{-\frac{\chi}{2}}\left(
\beta^\prime - \alpha
\right)^2
+q^2r^2fe^{-\frac{\chi}{2}}\psi^2\alpha^2
+k_a^2 q^2 e^{-\frac{\chi}{2}} \psi^2 \beta^2
\right.\nonumber\\
&&\left. \hspace{1.9cm}
+ r^2fe^{-\frac{\chi}{2}}\delta\psi^{\prime 2}
+ k_a^2e^{-\frac{\chi}{2}}\delta\psi^2
+ r^2 e^{-\frac{\chi}{2}} m^2 \delta\psi^2
\right.\nonumber\\
&&\left. \hspace{1.9cm}
- q^2r^2e^{-\frac{\chi}{2}}\phi^2\delta\psi^2
- q^2r^2e^{-\frac{\chi}{2}}\psi^2\left(
\lambda^2 + 4\frac{\phi}{\psi}\delta\psi \lambda
+ \frac{P_\alpha^\prime - P_\beta}{q^2r^2e^{-\frac{\chi}{2}}\psi^2}
\lambda
\right)
~\right]\,.
\label{hamiltonian2}
\end{eqnarray}
At first sight, the system seems to be unstable because of the term,
$- q^2r^2e^{-\frac{\chi}{2}}\phi^2\delta\psi^2$ in the last line of
 Eq.~(\ref{hamiltonian2}) in the limit of the Reissner-Nordstr\"{o}m AdS 
$\psi\rightarrow 0$.
However, $\psi $ is not actually zero. 
We can eliminate the unphysical variable $\lambda$ by completing the 
square for $\lambda$ or alternatively by using the 
equation of motion for $\lambda$
\begin{eqnarray}
 \lambda = - 2\frac{\phi}{\psi} \delta \psi
  - \frac{P^\prime_\alpha - P_\beta}{2q^2r^2e^{-\frac{\chi}{2}}\psi^2} \ .
\end{eqnarray}
Then, the last line of Eq.~(\ref{hamiltonian2}) becomes
\begin{eqnarray}
+ 3q^2r^2e^{-\frac{\chi}{2}}\phi^2\delta\psi^2
+ 2\frac{\phi}{\psi}\left(
P^\prime_\alpha - P_\beta
\right)\delta\psi
+ \frac{\left(P^\prime_\alpha - P_\beta\right)^2}
{4q^2r^2e^{-\frac{\chi}{2}}\psi^2} \,.
\label{after}
\end{eqnarray}
The effect we have incorporated is merely the backreaction of the gauge field.
Notice that the constraint equation, which is derived by the variation with
respect to $\lambda$, 
\begin{eqnarray}
P^\prime_\alpha - P_\beta = 0  
\end{eqnarray}
holds in the limit $\psi\rightarrow 0$. This constraint equation
is nothing but the Gauss law for the gauge field. This appears
because the limit  $\psi\rightarrow 0$ recovers the gauge invariance
in the gauge field system. 
Therefore, in the vicinity of the critical point, we can ignore last two terms
in Eq.(\ref{after}).  
Thus, when we take the limit $\psi\rightarrow 0$ from the side of the
superconducting phase, the Hamiltonian reduces to
\begin{eqnarray}
H&=& \int drd^2k_a \left[~
 \frac{1}{2r^2}e^{-\frac{\chi}{2}}P_\alpha^2
+ \frac{f}{2k_a^2}e^{-\frac{\chi}{2}}P_\beta^2
+ \frac{f}{4r^2}e^{-\frac{\chi}{2}}P_{\delta\psi}^2
+\frac{k_a^2 f}{2}e^{-\frac{\chi}{2}}\left(
\beta^\prime - \alpha
\right)^2
\right.\nonumber\\
&&\left. \hspace{1.9cm}
+ r^2fe^{-\frac{\chi}{2}}\delta\psi^{\prime 2}
+ k_a^2e^{-\frac{\chi}{2}}\delta\psi^2
+ r^2 e^{-\frac{\chi}{2}} m^2 \delta\psi^2
+ 3q^2r^2e^{-\frac{\chi}{2}}\phi^2\delta\psi^2
~\right] \ . 
\end{eqnarray}
Now, it is clear that the Hamiltonian is positive definite except for 
the mass term. Moreover, the effective mass squared
$m^2_{\rm eff} = m^2 +3q^2\phi^2$ is always above the BF bound. 
Thus, we find the system is going to be stable after the phase transition.
It should be stressed that the flip of the sign of the term
proportional to $\phi^2 \delta \psi^2$ is possible only in the condensed
phase with $\psi \neq 0$ everywhere. In fact, there is an (infinite) set of solutions
for $\psi$ which are regular on the horizon and satisfy the required boundary
condition at infinity.
 Based on our analysis, we believe that the solutions with nodes 
 are likely to be unstable and only the lowest solution with no nodes is stable. 

We can argue the stability of the superconducting phase far from
the critical point as follows.
From the point of view of the gravity dual, the phase transition
is the bifurcation of the solutions at the phase transition point
from where two branches are developed. The normal branch is apparently
unstable. What we have shown is the stability of the other branch
in the vicinity of the bifurcation point. However, from the continuity,
this is sufficient to prove the stability of the superconducting phase
provided that there exists no other bifurcation point on this branch.
Indeed, it is difficult to imagine a new bifurcation point
because the sign flip of the term
proportional to $\phi^2 \delta \psi^2$ persists as long as the scalar 
field has
the expectation value. Admittedly, this argument does not give a proof of 
stability far away from the critical point.

\section{Conclusion}

We have studied the dynamical stability of superconductors
using the ADM formalism. 
First, we proved the stability of the vector sector by explicitly
constructing the positive definite conserved Hamiltonian.
We have also obtained the Hamiltonian for the scalar sector.
In the normal phase, the action decoupled into two parts.
One is that of the Reissner-Nordstr\"{o}m-AdS black hole and the other is that of
the charged scalar field in the Reissner-Nordstr\"{o}m-AdS background.
It  is well known that the former system is stable. The stability of the latter
system depends on the effective mass of the scalar field.
Since the effective mass becomes more negative as the gauge potential becomes large,
the phase transition occurs at the critical temperature. 
And, below the critical temperature, the Reissner-Nordstr\"{o}m-AdS
black holes are unstable.
Numerically, we know the scalar hair is developed below $T_c$.
For this case, the structure of the constraint for $\lambda$ changes.
After eliminating the unphysical variable $\lambda$, we found the effective
mass became above the BF bound. In particular, near the critical temperature 
$\psi \rightarrow 0$, we have decoupled equations for the charged scalar field,
which is now stable.
Thus, we have revealed a mechanism for the stabilization of 
superconducting phase in the scalar sector. 
The result is remarkable because the stability near the critical point
 is guaranteed by the non-gravitational effect. In fact, 
the effect of the gauge field changes
the signature of the effective additional mass when the charged
scalar field has condensation.
In our argument, we have assumed the solution for the scalar field has no node.
Therefore, other solutions with nodes
are expected to be unstable. Admittedly, we have not been able to
prove the stability of superconductors far away from the critical point. 
However, the fact that the effective mass squared remains to be above the BF bound
 in the superconducting phase suggests the stability of holographic superconductors.
It should be stressed that the stabilization mechanism we found 
originates from the structure of the minimal gauge coupling, hence the result
 applies to holographic superconductors in any dimensions. 

Our analysis can be extended to various holographic  superconductor models.
For example, it is possible to extend our analysis to the Gauss-Bonnet
superconductors~\cite{Gregory:2009fj,Pan:2009xa,Brihaye:2010mr,
Liu:2010bq,Cai:2010cv}. 
In that case, again, we can use the results
in black hole perturbations~\cite{Dotti:2004sh,Takahashi:2009dz}. 

More importantly, we need to study the dynamical stability of
holographic superconductors embedded into string 
theory~\cite{Gubser:2009qm,Gauntlett:2009dn,Gauntlett:2009bh}. 
However, it requires a more sophisticated method for the analysis.

\begin{acknowledgements}
SK would like to thank TAP members in Kyoto University for warm
hospitality. Most of this work was done while SK was visiting Kyoto
University supported by JSPS Grant-in-Aid for Scientific Research 
on Innovative Area No.21111006.
SK is supported by an STFC rolling grant.
JS is supported by  the
Grant-in-Aid for  Scientific Research Fund of the Ministry of 
Education, Science and Culture of Japan No.22540274, the Grant-in-Aid
for Scientific Research (A) (No. 22244030), the
Grant-in-Aid for  Scientific Research on Innovative Area No.21111006
and the Grant-in-Aid for the Global COE Program 
``The Next Generation of Physics, Spun from Universality and Emergence". 
\end{acknowledgements}

\end{document}